\documentstyle{aipproc}

\begin{document}
\tightenlines
\title{Pion Scattering Revisited}
\author{M. Ruiz-Altaba, J.L. Lucio$^*$ and M. Napsuciale$^*$}

\address{Departamento de F\'{\i}sica Te\'orica, 
Instituto de F\'{\i}sica \\
Universidad Nacional Aut\'onoma de M\'exico, 
A.P. 20-364, 
01000 M\'exico, D.F.\\
$\quad$ \\
$^*$Instituto de F\'\i sica, Universidad de Guanajuato\\
Loma del Bosque 103, 37160 Le\'on, Guanajuato, M\'exico}
\maketitle

\begin{abstract}
Chiral Ward identities lead to consistent accounting for the 
$\sigma$'s 
width in the linear sigma model's Feynman rules. Reanalysis of 
pion scattering data at threshold imply a mass  for  the $\sigma$ 
of  $  600{+ 200\atop - 100} $MeV.
\end{abstract}

This short talk (by M.R-A) reviews our recent work on the linear 
sigma model \cite{nos,mau}, where full references can be found.  
At low  energies, chiral perturbation theory is supposed to yield 
good agreement with strong interaction data. Unfortunately, 
chiral perturbation theory  gives rather poor  results on the 
scattering lengths of pion-pion scattering, which are  relevant 
experimental quantities in the limit of zero momentum, that is to 
say, where chiral perturbation theory should work best. 

A missing ingredient in the description at low energies of strong 
interactions is the $\sigma$ field, in addition to the Goldstone 
bosons of chiral symmetry (the pions). A wide scalar resonance in 
the vicinity of 600 MeV exists, and can be identified naturally 
with the $\sigma$ particle of the original {\sl linear} 
$\sigma$--model.

What are the phenomenological consequences of the linear 
$\sigma$--model in $\pi\pi\to\pi\pi$ scattering at very low 
energies?  The sole guiding principle is chiral symmetry, whose 
Ward identities allow us modify the various vertices to take into 
account the large width of the $\sigma$ resonance. 

The  chiral symmetry breaking giving mass to the pions is soft, 
so that when we   include the width $\Gamma_\sigma$ of the 
$\sigma$ in its propagator, we can exploit the chiral Ward 
identities to modify the vertices accordingly. The chiral Ward 
identities are  satisfied by the resulting lagrangian (with 
parameters $m_\pi$, $f_\pi$, $m_\sigma$), from which we compute 
the amplitudes in the various isospin and angular momentum 
channels of experimental relevance. We  use the expression for 
$\Gamma_\sigma$ from the decay $\sigma\to\pi\pi$ to perform a 
simple and succesful one--parameter ($m_\sigma$) fit to data.

The field $\sigma$ is very unstable: its tree--level width is 
$$\Gamma\left(\sigma\to \pi\pi\right)=
{3m_\sigma^3 \over32\pi f_\pi^2} 
(1-\varepsilon)^2\sqrt{1-4\epsilon} $$
where we have introduced the convenient shorthand 
$\varepsilon=(m_\pi/m_\sigma)^2$.   In strict analogy with the 
Higgs field in the standard model, the $\sigma$ width 
$\Gamma_\sigma$ grows very fast with its mass: $\Gamma_\sigma(350 
)= 65$, $\Gamma_\sigma(500 )= 310$, $\Gamma_\sigma(650 )= 785$, 
all in MeV. The effect of the width of the $\sigma$ field is to 
modify its propagator from the usual $i\left( q^2-m_\sigma^2 
\right)^{-1}$ to
$\Delta_\sigma(q)= i\left( q^2-m_\sigma^2 +i\Gamma_\sigma 
m_\sigma \, \theta( q^2-4m_\pi^2) \right)^{-1}$,
where the step function ensures that the imaginary piece in the 
denominator appears only when the momentum of the propagator is  
above the kinematical threshold for $\sigma$ decay. 

Thus, in the physical process of $\pi\pi\to\pi\pi$ scattering, 
which we shall consider shortly, the propagator of the $\sigma$  
picks up the correction due to the width only in the 
$s$--channel, not in the $u$-- nor the $t$--channels.

The crucial point is that, in the linear $\sigma$ model, chiral 
symmetry is responsible for important cancellations which imply, 
notably, that the pion coupling is always derivative in the limit 
of soft pion momenta. 
Enforcing  the chiral Ward identities on the vertices of the 
lagrangian  implies that the latter pick up modifications related 
to the width $\Gamma_\sigma$. These vertex corrections depend on 
the kinematical variables (the incoming momenta) in a particular 
way, dictated by chiral symmetry. For instance, the $\sigma 
\pi^i\pi^j$ Feynman rule reads now $$ V_{\sigma \pi^i\pi^j}= 
{-i\over f_\pi} \delta^{ij} \left( m_\sigma^2 -m_\pi^2 
-i\Gamma_\sigma m_\sigma \theta(q^2-4m_\pi^2)\right)$$
where $q^\mu$ is the momentum of the $\sigma$.

We find also
$$V_{\pi^i\pi^j\sigma\sigma} = V_{\sigma\sigma \sigma} 
\Delta_\sigma(p_j)   V_{\sigma\pi ^i\pi^j} $$ where $p_j$ is the 
momentum of a pion, so that $p_j^2 =m_\pi^2$ if it is on--shell. 
This equation {\sl defines} the $\pi\pi\sigma\sigma$ vertex. 
Similarly, the chiral Ward identity satisfied by the $\pi^4$ 
Feynman rule is 
$$ V_{\pi^i\pi^j\pi^k\pi^\ell} = V_{\pi^k\pi^\ell \sigma} 
\Delta_\sigma(p_j)   V_{\sigma\pi ^i\pi^j} +V_{\pi^i\pi^k \sigma} 
\Delta_\sigma(p_k)   V_{\sigma\pi ^j\pi^\ell} +V_{\pi^i\pi^\ell 
\sigma} \Delta_\sigma(p_\ell)   V_{\sigma\pi ^j\pi^k}$$
Obviously, these relations hold at tree level before chiral 
symmetry breaking, that is to say, when $m_\pi=0$, and also 
$\Gamma_\sigma=0$. Powefully, they also hold when $m_\pi\not=0$ 
and/or when $\Gamma_\sigma\ne0$, to all orders in perturbation 
theory. This can be proved easily using the enormous advantage 
that the linear sigma model is a well--defined (renormalizable) 
field theory.  

Since the vertex modifications ensure the preservation of exact 
chiral Ward identities, they  also guarantee, for instance, that 
the pion couplings remain derivative as they should. 

To illustrate the power of this implementation of chiral 
symmetry, we  evaluate, at tree level, the amplitude for $\pi\pi$ 
scattering. Clearly, we do not expect the result to be the 
perfect answer, since the only resonance we will take into 
account is the $\sigma$. In particular,  not taking into account 
the vector meson $\vec{\rho}^\mu$ is a rather bad approximation 
in the $I=1$, $\ell=1$ amplitude. Nevertheless, our results are 
in better agreement with experimental data than those of chiral 
perturbation theory. Let us emphasize that the kinematical region 
where we compare both predictions, namely at very low momenta, is 
precisely where chiral perturbation theory should be exact. This 
lends further support to the real existence 
of $\sigma$ as a strong resonance. 

At tree level, four diagrams contribute to $\pi\pi\to\pi\pi$:  
the four--pion contact term, and the exchange of a $\sigma$ in 
the three $s$, $t$ and $u$ channels. 
Due to the structure of the Feynman rules dictated by chiral Ward 
identities, the width $\Gamma_\sigma$ contributes, in the Born 
approximation, only to $T^{(0)}_0$.  
 
The experimental knowledge of pion scattering near threshold is 
rather poor. The relatively badly measured scattering lengths and 
ranges are  $a_0^{(0)}$, $b_0^{(0)}$, $a_0^{(2)}$, $b_0^{(2)}$, 
$a_1^{(1)}$, $a_2^{(0)}$ and  $a_2^{(2)}$, These seven numbers 
come out of our computation with only $m_\sigma $  as a free 
parameter.

An overall fit to these seven numbers gives  $m_\sigma=700 {+ 800 
\atop -150}$MeV. The   $\chi^2$  distribution is very flat 
towards increasing values of $m_\sigma$;     $m_\sigma\ge$550 Mev 
is the only useful information. 

Of the seven numbers, if we eliminate the worst one ($a_1^{(1)}$ 
(presumably under strong influence from $\rho$ exchange, which we 
do not take into account), then the fit improves and it yields 
$m_\sigma= 590 { +220 \atop -90}$MeV. Nicely,  the fit to only 
the scalar isoscalar values gives  
$m_\sigma= 525 { +85 \atop -45}$MeV. 

Overall, one may conclude that the data are consistent with a 
linear sigma resonance provided its mass is around 600 MeV (and 
thus its width also around 600 MeV). The errors on these numbers, 
from the pion data available, are substantial. 

Although the low--energy moments $a_\ell^{(I)}$ and 
$b_\ell^{(I)}$ are the relevant quantities for us, what is 
actually measured is a momentum--dependent phase shift, which can 
be split in various isospin and angular momentum channels. 
From the analysis of the data available, we   fit    
$m_\sigma=550 {+450 \atop -80}$MeV. Again   the error on the 
heavy side is huge: the $\chi^2$ distribution is very flat with 
increasing  $m_\sigma$.

Exact unitarity is achieved iff
$$ {\rm Im}\, T_\ell^{(I)} = \sqrt{ s-4m_\pi^2 \over s}  \left|  
T_\ell^{(I)} \right|^2 $$
from which the optical theorem can be derived.  Since there are 
many other resonances in nature heavier than the $\sigma$, we 
should not worry much about possible unitarity violations at high 
momenta (say, above 1 GeV).
It turns out that there is no problem with unitarity at center of 
mass   momenta lower than the 400 MeV. Unfortunately, unitarity 
does not constrain $m_\sigma$ from above in any meaningful way.

We have enhanced the linear sigma model by enforcing chiral Ward 
identities which take into account the (large) sigma width. 
We have found that low energy pion scattering data supports the 
existence of a wide $\sigma$ field with mass around 600 MeV 
(actually $m_\sigma= 590 { +220 \atop -90}$MeV), provided we 
exclude the datum in the vector isovector channel. The advantage 
of keeping the $\sigma$ as a true resonance in the effective low 
energy theory of strong interactions is not only that its 
inclusion simulates more or less the results of chiral 
perturbation theory to one loop,  but also, more crucially, that  
 this opens the door to more industrious analyses of the whole 
scalar spectrum, including glueballs.

\vskip.2cm{\bf Acknowledgements}. 
This work was supported in part by CONACYT through projects 
3979P-E9608, 25504-E,  and C\'atedra Patrimonial II de Apoyo a 
los Estados, and by DGAPA--UNAM through IN103997.

\end{document}